\documentclass[intlimits,twoside,a4paper]{article}
\usepackage[cp1251]{inputenc}

\usepackage{cmpj3}

\issue{2018}{21}{2}{23802}
\doinumber{10.5488/CMP.21.23802}

\title[Thermodynamics of primitive model electrolytes]
{Thermodynamics of primitive model electrolytes in the symmetric and modified Poisson-Boltzmann theories. A comparative study with Monte Carlo simulations}

\author[A.O. Qui\~{n}ones, L.B. Bhuiyan, C.W. Outhwaite]{A.O.~Qui\~{n}ones\refaddr{label1},
L.B.~Bhuiyan\refaddr{label1}, C.W.~Outhwaite\refaddr{label2}}

\addresses{
\addr{label1}Laboratory of Theoretical Physics, Department of
Physics, University of Puerto Rico, 17 Avenida Universidad, STE 1701, San Juan, Puerto
Rico 00925-2537, USA
\addr{label2}Department of Applied Mathematics, University of Sheffield,
Sheffield S3 7RH, UK
}

\date{Received April 23, 2018}

\begin{document}

\maketitle
\begin{abstract}

Osmotic coefficients, individual and mean activity coefficients of
primitive model electrolyte solutions are computed at different molar concentrations
using the symmetric Poisson-Boltzmann and modified Poisson-Boltz\-mann theories.
The theoretical results are compared with an extensive series of Monte Carlo
simulation data obtained by Abbas et al. [Fluid Phase Equilib., 2007, 
\textbf{260}, 233; J. Phys. Chem. B, 2009, \textbf{113}, 5905]. The agreement 
between modified Poisson-Boltzmann predictions with the ``\emph{exact}'' simulation 
results is almost quantitative for monovalent salts, while being semi-quantitative 
or better for higher and multivalent salts. The symmetric Poisson-Boltzmann results, 
on the other hand, are very good for monovalent systems but tend to deviate at 
higher concentrations and/or for multi-valent systems. Some recent experimental 
values for activity coefficients of HCl solution (individual and mean activities) 
and NaCl solution (mean activity only) have also been compared with the symmetric 
and modified Poisson-Boltzmann theories, and with the Monte Carlo simulations.

\pacs 82.45.Fk, 61.20.Qg, 82.45.Gj
\end{abstract}
\section{Introduction}

Designing many of the important industrial chemical processes involving
aqueous electrolytes requires an understanding of the thermodynamics
of these systems. Ion-ion or ion-surface charge interactions of electrolyte solutions
in water are fundamental processes also in areas such as biology and physical chemistry
(see for example, references \cite{harned,robinson,pitzer1,pitzer2,loehe,levine,kalyuzhnyi,
levin,chersty,may}). These interactions relate to the activity of water and the individual
activities of the ions in solution. Two of the more important properties of such Coulomb fluids,
which describe the thermodynamics, are the osmotic coefficient $\phi $ and the activity coefficient
$\gamma $. These are experimentally measurable quantities \cite{robinson}
that are widely used in physical chemistry and electrochemistry to quantify the deviation
of experimental measurements from theoretically predicted ideal cases.

Theoretically, the classical approach to estimating the osmotic and activity coefficients
is based on the Debye-H\"{u}ckel (DH) theory \cite{debye}. However, due to its mean-field nature
and, more importantly, due to its linear character, the theory is somewhat restricted in its applications.
The Debye-H\"{u}ckel limiting law (DHLL), however, provides useful limiting expressions for
many physical quantities of interest \cite{mcquarrie}. A lasting legacy of the DH theory
is the underlying physical model that it portrays, that is, the point ions moving in a dielectric
continuum. By imparting a size to the ions we arrive at the primitive model (PM) of electrolytes, viz.,
charged hard spheres in a dielectric continuum. If the sizes of the spheres are the same,
then we have a restricted primitive model (RPM). The RPM and the PM have been extensively
used over the years in formal statistical mechanical theories to describe the structure and
thermodynamics of charged fluids in the bulk and near charged interfaces. Mention can be
made of liquid structure integral equations such as the hypernetted chain
\cite{hansen} and the mean spherical approximation \cite{hansen,blum,blum2},
and potential based approaches, viz., the symmetric Poisson-Boltzmann (SPB)
\cite{outh1,martinez,molero,outh2} and modified Poisson-Boltzmann (MPB)
\cite{outh2,outh3,outh4,ulloa} theories. Parallel numerical simulations, for example,
using Monte Carlo (MC) and Molecular Dynamics simulations \cite{vlachy} have been valuable
in theoretical developments.

A few years ago, Abbas et al. \cite{abbas1,abbas2} made extensive MC simulations for a series
of symmetric (in ion size and valence) and asymmetric PM electrolytes and reported their osmotic
coefficient, single ion and mean activity coefficient results. In total, they treated 104
electrolyte systems covering a wide range of concentration and different valence combinations.
The object of their work was to see how valid the PM was in representing the experimental
results of salt solutions. Ionic activity and osmotic coefficients were calculated for
1$^{+}$:1$^{-}$,2$^{+}$:1$^{-}$, and 3$^{+}$:1$^{-}$ electrolytes and best fitting ionic
radii determined for describing the experimental results. In this paper, we focus on obtaining
the corresponding SPB and MPB results for these systems \cite{abbas1,abbas2}, and compare the
theoretical predictions against the \emph{exact} MC data.

Experimental measurements of osmotic and mean activity coefficients of electrolytes
also abound in the literature \cite{robinson,pitzer1,pitzer2}. These two quantities are related
by the famous Gibbs-Duhem equation of physical chemistry \cite{harned}, which has now been extended
to multicomponent fluids \cite{vlachy2}. The advent of sophisticated experimental techniques over the
past two decades has made determination of activity of individual ionic species in electrolytes
increasingly feasible \cite{vera}. Here, we will also make comparisons of the SPB and MPB predicted
activity coefficients with the experimental results for HCl due to Sakaida and Kakuichi \cite{sakaida}
and for NaCl taken from the literature \cite{robinson}.

\section{Model and Methods}

As suggested in the Introduction we have treated here the PM of the electrolyte,
that is, the constituent ions are mimicked by rigid spheres of arbitrary radii with an embedded
charge of arbitrary valence at the centre of each sphere. The charged hard spheres move in a
continuum solvent characterized by a single parameter, viz., a relative permittivity $\epsilon _\text{r}$.
The model is the one used in the MC studies of Abbas et al. \cite{abbas1,abbas2}.

The pair interaction potential in the Hamiltonian is given by
\begin{equation}
 u_{ij}(r)=\left\{
\begin{array}{cc}
 \infty & r<(r_{i}+r_{j}) \\
\frac{1}{4\piup \epsilon_{0}\epsilon_\text{r}}\frac{e^{2} Z_i Z_j}{r} & r>(r_{i}+r_{j})
\end{array}
\right.,
\end{equation}

\noindent where $r_{s}$ and $Z_{s}$ are the radius and valence of ion species $s$, $e$ is the
proton charge, $r$ is the separation between the centers of ions $i$ and $j$, and $\epsilon_{0}$
the vacuum permittivity. In a special case when the ions are of the same size, that is, $r_{i} = r_{j}$,
we have the RPM.

The above PM (or the RPM) was treated by the symmetric Poisson-Boltzmann and the modified
Poisson-Boltzmann theories. The development and the formulation of the SPB and the MPB have been
detailed elsewhere in the literature (see references
\cite{outh1,martinez,molero,outh2}
for the SPB and references
 \cite{outh2,outh3,outh4} for the MPB) and will not be repeated here.

\section{Results}

The SPB and MPB equations were solved numerically using a quasi-linearization
iteration scheme \cite{bellman}. This is a well tested, robust method, which has
been successfully utilized in earlier works involving the SPB and MPB \cite{outh1,outh3}.

The results presented in this paper correspond to  1:1, 2:1, and 3:1
valence electrolytes at temperature $T =298$ K  in
a water-like solvent with a dielectric constant $\epsilon_\text{r}=78.5$, and for a wide range
of concentration. These values are consistent with some experiments we have compared our results
with \cite{robinson,sakaida} and the MC simulations of Abbas et al. \cite{abbas1,abbas2}. The ion sizes
used are also from their simulations, and although ions are generally not spherical,
the ion sizes represent optimized values against experimental osmotic
coefficients and mean activity coefficients.

\subsection{Comparison with experiments}

We begin this discussion by comparing the SPB and MPB activity coefficient results
against some experimental data for HCl and NaCl. There is some debate in the literature over the
experimental determination of individual ion activities \cite{vera,fraenkel}. Wilczek-Vera and Vera
\cite{vera} have compared some experimental results for different salt systems and have found, in many
situations, a reasonable agreement between the results for the salts using different techniques.

\subsubsection{HCl}

\begin{figure}[!b]
\centering
\begin{minipage}{0.49\textwidth}
\begin{center}
\includegraphics[width=0.99\textwidth]{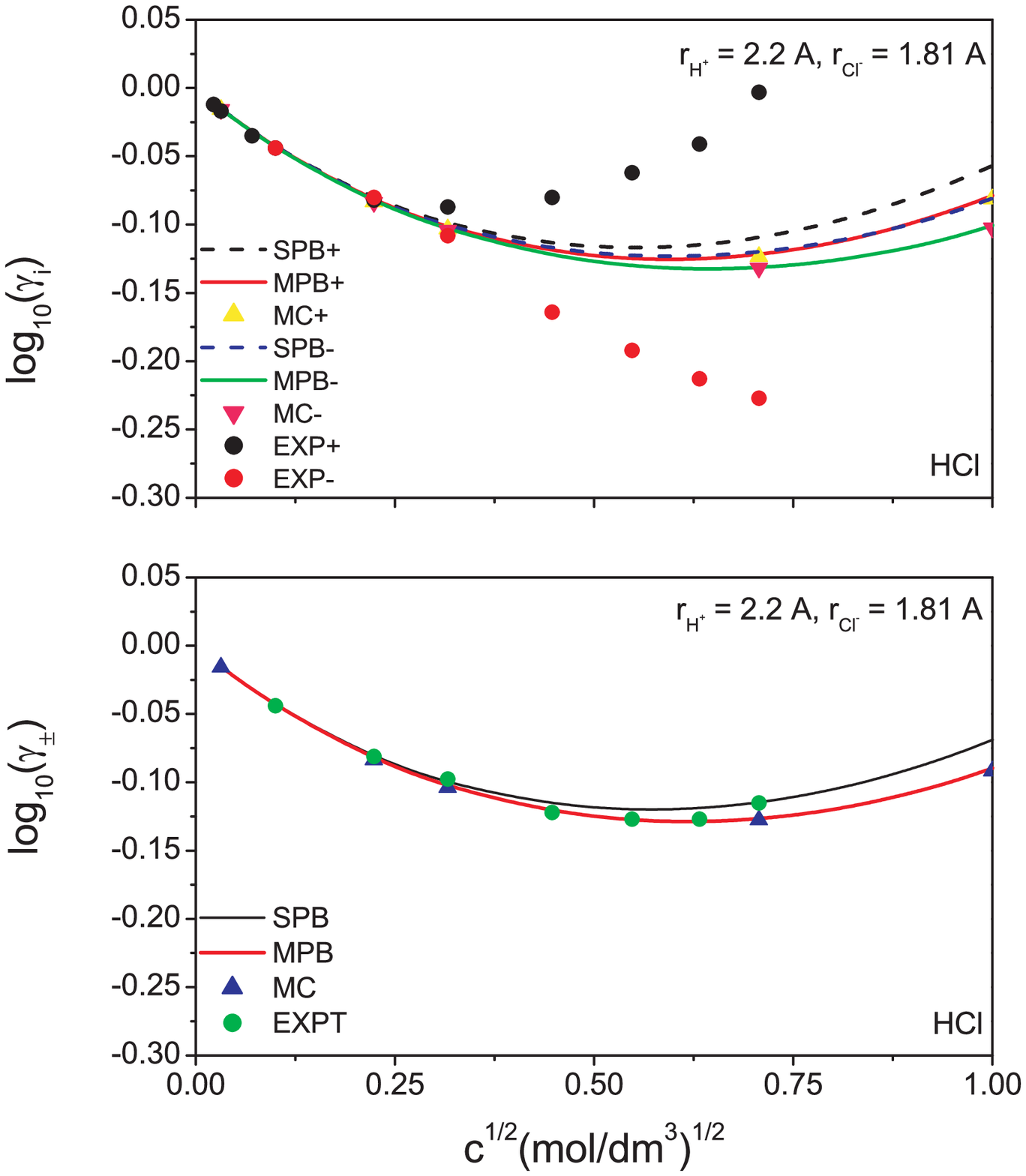}
\end{center}
\end{minipage}
\begin{minipage}{0.49\textwidth}
\begin{center}
\includegraphics[width=0.99\textwidth]{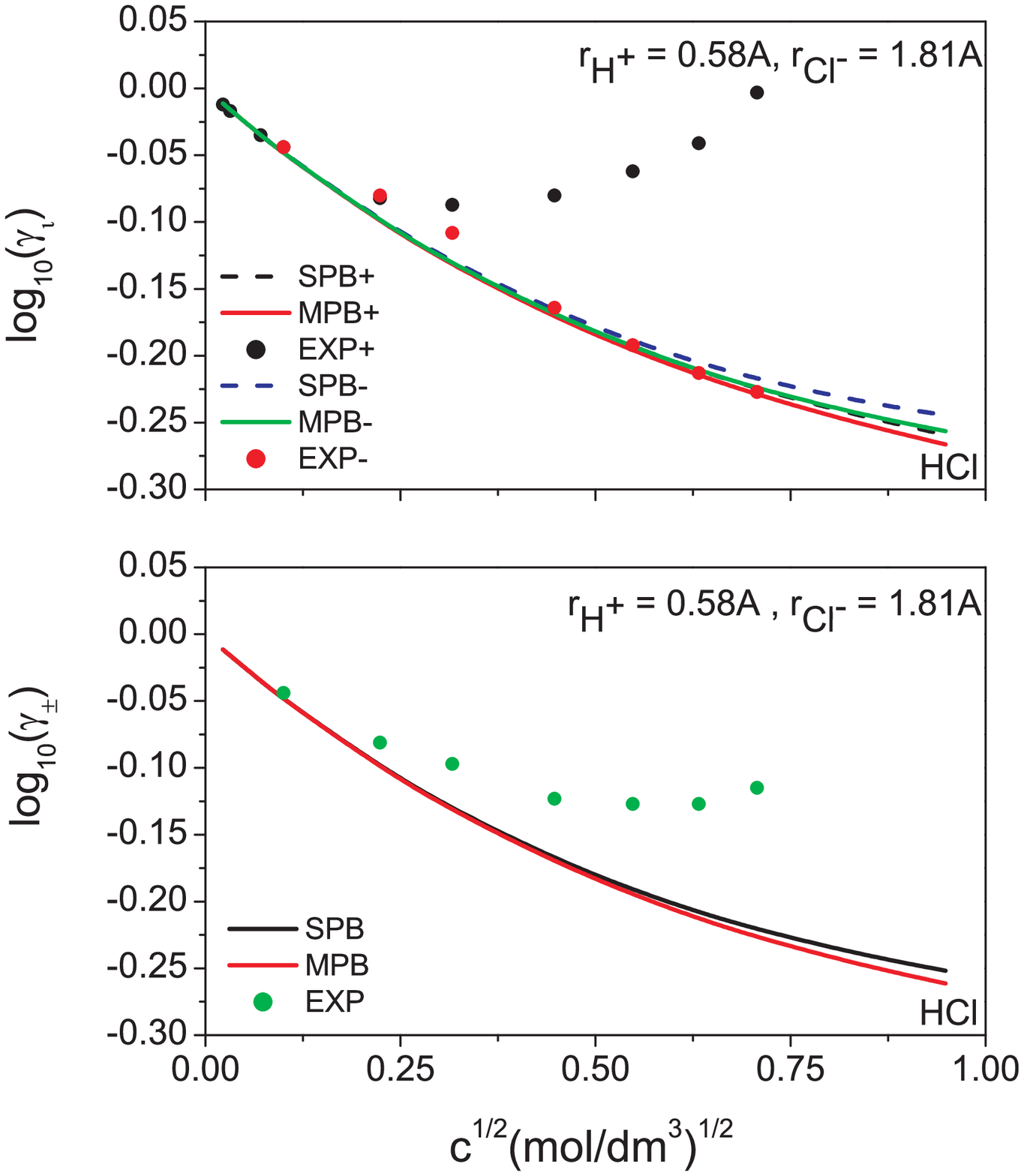}
\end{center}
\end{minipage}
\begin{minipage}{0.49\textwidth}
\begin{center}
\caption{(Colour online) Experimental, theoretical (SPB and MPB), and MC decadic logarithm
		of individual and mean activity coefficients of HCl solution as functions of electrolyte concentration.
		In the SPB and MPB calculations, the ionic radii are $r_{\text{H}^{+}} = 2.20 \times 10^{-10}$ m,
		$r_{\text{Cl}^{-}} = 1.81 \times 10^{-10}$ m, and are taken from the MC simulation data of references
		\cite{abbas1,abbas2}. The experimental data are from \cite{sakaida}.}
	\label{figure-1}
\end{center}
\end{minipage}
\begin{minipage}{0.49\textwidth}
\begin{center}
\caption{(Colour online)  Experimental and theoretical (SPB and MPB) decadic logarithm of
		individual and mean activity coefficients of HCl solution as functions of electrolyte concentration.
		In the SPB and MPB calculations, the ionic radii are $r_{\text{H}^{+}} = 0.58 \times 10^{-10}$ m,
		$r_{\text{Cl}^{-}} = 1.81 \times 10^{-10}$ m, and are taken from the SiS theory of Fraenkel \cite{fraenkel1}.}
	\label{figure-2}
\end{center}
\end{minipage}
\end{figure}


Sakaida and Kakiuchi have measured the individual activity coefficients of HCl
(and hence the mean activity coefficient) at room temperature using an ionic liquid salt
bridge \cite{sakaida}. We have calculated the corresponding SPB and MPB results for this
system using the H$^{+}$, Cl$^{-}$ radii taken from (a) the Abbas et al. \cite{abbas1,abbas2}
MC simulation data,viz., $r_{\text{H}^{+}} = 2.20 \times 10^{-10}$ m, $r_{\text{Cl}^{-}} = 1.81 \times 10^{-10}$ m,
and (b) from Fraenkel's \cite{fraenkel1} theory, $r_{\text{H}^{+}} = 0.58 \times 10^{-10}$ m,
$r_{\text{Cl}^{-}} = 1.81 \times 10^{-10}$ m. The experimental data are in ``molal'' scale, whereas
the MC and theoretical data are in ``molar'' scale. For consistency, we have converted
the experimental data into ``molar'' units. However, the difference between the two
sets of data are negligible since the range of concentration used in the experiment is small.
The experimental and theoretical results for the individual and mean activity coefficients
for this system at various concentrations are shown in figures \ref{figure-1} and \ref{figure-2}, respectively. Following
literature custom we have plotted the logarithms of the activity coefficients as function of
the square root of the salt concentration. In each figure, the top panel shows the individual
activity coefficients, while the bottom panel gives the mean activity coefficients.

In figure \ref{figure-1} (top panel) the experimental, theoretical, and simulation individual activity
coefficient data are close together up to about $c\sim 0.07$~mol/dm$^{3}$. Indeed, in the limit
of low concentration, they tend to the same limiting value consistent with the DHLL. However, as the
concentration increases beyond $c\sim 0.07$~mol/dm$^{3}$, the simulation and the theoretical
activities deviate from the experimental results, especially for the cationic activity
coefficient. It is noted that for the range of concentration plotted, the SPB and MPB results follow
the simulations rather well. Although the MPB difference is almost quantitative with the MC, the SPB shows
deviations at higher concentrations, which can be attributed to the neglect of ionic correlations
in the mean-field theory. Such correlations become more significant for dense solutions. In contrast
to the situation in the top panel, the experimental, the MC, and the theoretical mean activity
coefficients in the bottom panel remain relatively close to each other throughout, which suggests some
error cancellations for the MC and the theories.


The MC data are conventionally taken to be ``exact'' for a given physical model. These results
suggest that the PM probably does not lead to an adequate physical description of HCl at high
concentrations. Fraenkel \cite{fraenkel1,fraenkel2} took this idea further and developed the
Smaller ion Shell (SiS) theory, which is based on an improved DH type theory with different ion sizes, and
gives a better representation of the HCl experimental individual activity coefficients \cite{sakaida}.
The SPB and MPB activities using ionic radii from Fraenkel's theory \cite{fraenkel1} are given in figure~\ref{figure-2}. Although MC data are not available at these ionic radii, from the consistency of the simulation
data with the theories for all the 104 cases for which the MC data are available, it is a fair
conjecture that the SPB and MPB curves in this figure will be very close to the ``exact'' results
for the model. The relative behaviour of the theoretical curves relative to the experimental curves
are similar to that in figure \ref{figure-1}. At these radii too, the shortcomings of the PM are clearly seen.

\subsubsection{NaCl}

\begin{figure}[!b]
	\centerline{\includegraphics[width=0.5\textwidth]{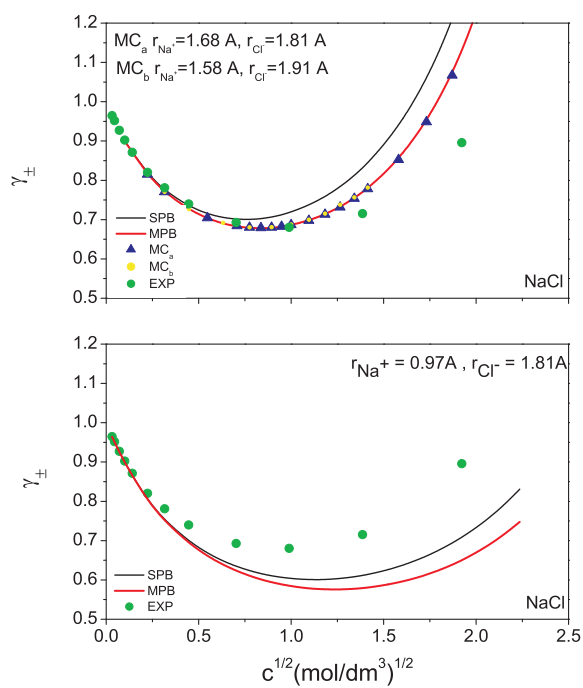}}
	\caption{(Colour online) Experimental and theoretical (SPB and MPB) mean
		activity coefficients of NaCl solution as functions of electrolyte concentration and for
		three sets of ionic radii, viz.,
		\emph{(i)} $r_{\text{Na}^{+}} = 0.97 \times 10^{-10}$ m, $r_{\text{Cl}^{-}} = 1.81 \times 10^{-10}$ m,
		\cite{fraenkel1} (lower panel),
		\emph{(ii)} $r_{\text{Na}^{+}} = 1.68 \times 10^{-10}$ m, $r_{\text{Cl}^{-}} = 1.81 \times 10^{-10}$~m
		\cite{abbas1,abbas2} (upper panel), and
		\emph{(iii)} $r_{\text{Na}^{+}} = 1.58 \times 10^{-10}$ m, $r_{\text{Cl}^{-}} = 1.91 \times 10^{-10}$ m
		\cite{abbas1,abbas2} (upper panel).}
	\label{figure-3}
\end{figure}

We have plotted the SPB and MPB results for the mean activity coefficients at
three sets of ionic parameters:
\emph{(i)} $r_{\text{Na}^{+}} = 0.97 \times 10^{-10}$ m, $r_{\text{Cl}^{-}} = 1.81 \times 10^{-10}$ m \cite{fraenkel1},
\emph{(ii)} $r_{\text{Na}^{+}} = 1.68 \times 10^{-10}$ m, $r_{\text{Cl}^{-}} = 1.81 \times 10^{-10}$ m \cite{abbas1,abbas2}, and
\emph{(iii)} $r_{\text{Na}^{+}} = 1.58 \times 10^{-10}$ m, $r_{\text{Cl}^{-}} = 1.91 \times 10^{-10}$ m \cite{abbas1,abbas2}.
The experimental values of the mean activity coefficients are from Robinson
and Stokes \cite{robinson}, and are given in terms of $\gamma _{\pm}$ as function of concentration,
hence the MC, SPB, and MPB plots of $\gamma _{\pm}$ for consistency.  The results are shown in figure \ref{figure-3}. At the Fraenkel ion-size parameters (lower panel), the theoretical mean activities are qualitative with the experiment, with no MC data being available at these parameters. However, in the upper panel of the figure, for both sets
of ionic radii, the MPB  mean activities are in excellent agreement with the MC data with the SPB showing
deviations (from the MC and MPB) as concentration increases. The theories and the simulations are again
qualitative and relatively closer to the experimental trends than with the Fraenkel parameters..

\subsection{Comparison with simulations}

\begin{figure}[!b]
	\centerline{\includegraphics[width=0.5\textwidth]{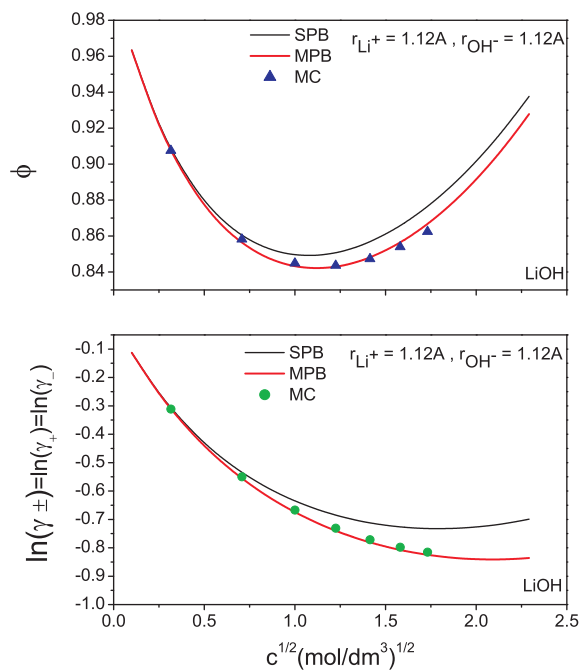}}
	\caption{(Colour online) MC and theoretical (SPB and MPB) osmotic coefficient (upper panel),
		and natural logarithm of the activity coefficients (individual and mean) (lower panel) for
		LiOH using RPM. The MC data are from references \cite{abbas1,abbas2}.}
	\label{figure-4}
\end{figure}

\begin{figure}[!t]
\centering
\begin{minipage}{0.49\textwidth}
\begin{center}
\includegraphics[width=0.99\textwidth]{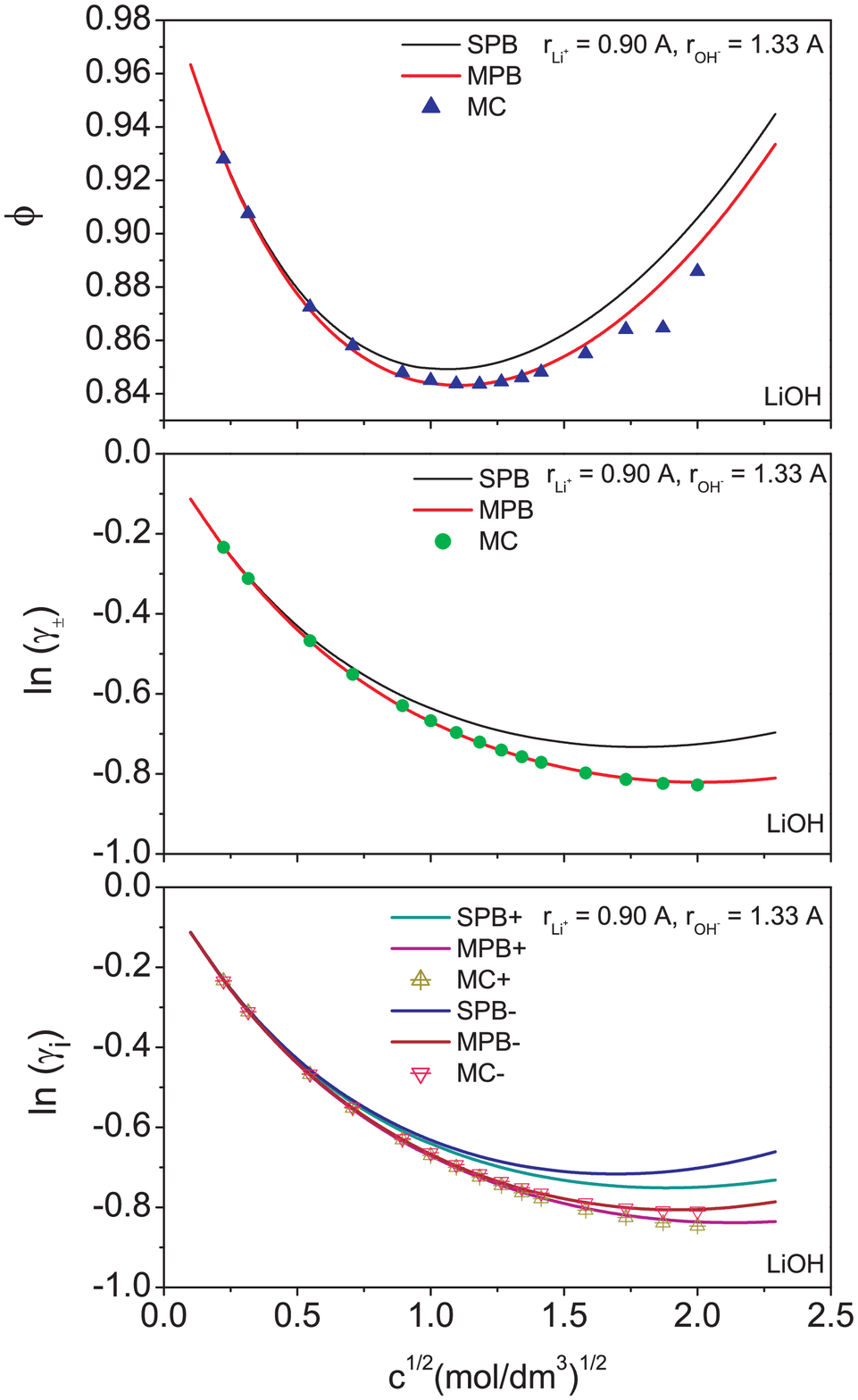}
\end{center}
\end{minipage}
\begin{minipage}{0.49\textwidth}
\begin{center}
\includegraphics[width=0.98\textwidth]{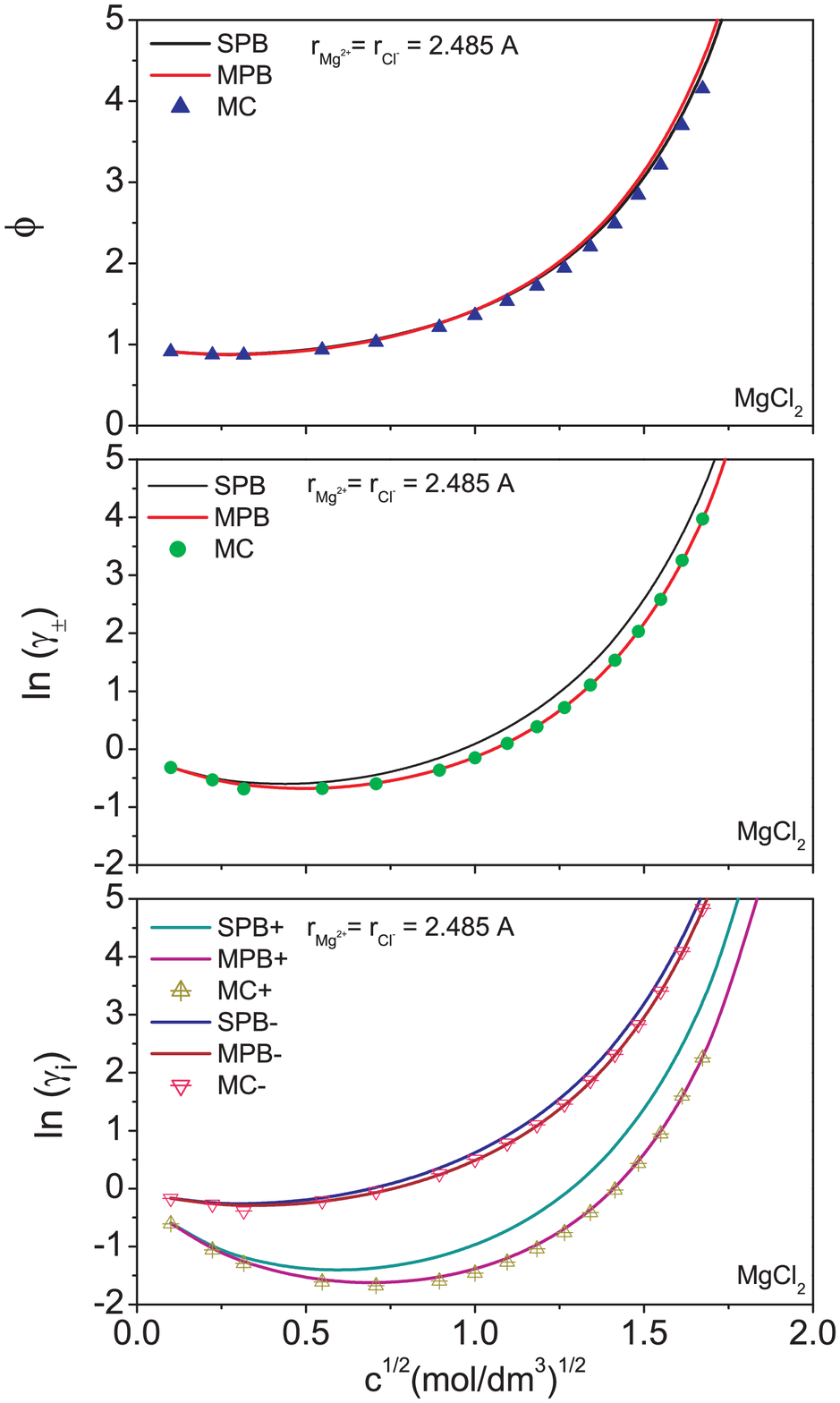}
\end{center}
\end{minipage}
\begin{minipage}{0.49\textwidth}
\begin{center}
\caption{(Colour online) MC and theoretical (SPB and MPB) osmotic coefficient (upper panel),
	natural logarithm of the mean activity coefficients (middle panel), and natural logarithm of the
	individual activity coefficients for LiOH using PM (lower panel). The MC data are from references
	\cite{abbas1,abbas2}.}
\label{figure-5}
\end{center}
\end{minipage}
\begin{minipage}{0.49\textwidth}
\begin{center}
\caption{(Colour online) MC and theoretical (SPB and MPB) osmotic coefficient (upper panel),
	natural logarithm of the mean activity coefficient (middle panel), and natural logarithm of the individual activity coefficients for MgCl$_{2}$ using RPM (lower panel). The MC data are from references \cite{abbas1, abbas2}.}
\label{figure-6}
\end{center}
\end{minipage}
\end{figure}


\begin{figure}[!t]
\centering
\begin{minipage}{0.48\textwidth}
\begin{center}
\includegraphics[width=0.99\textwidth]{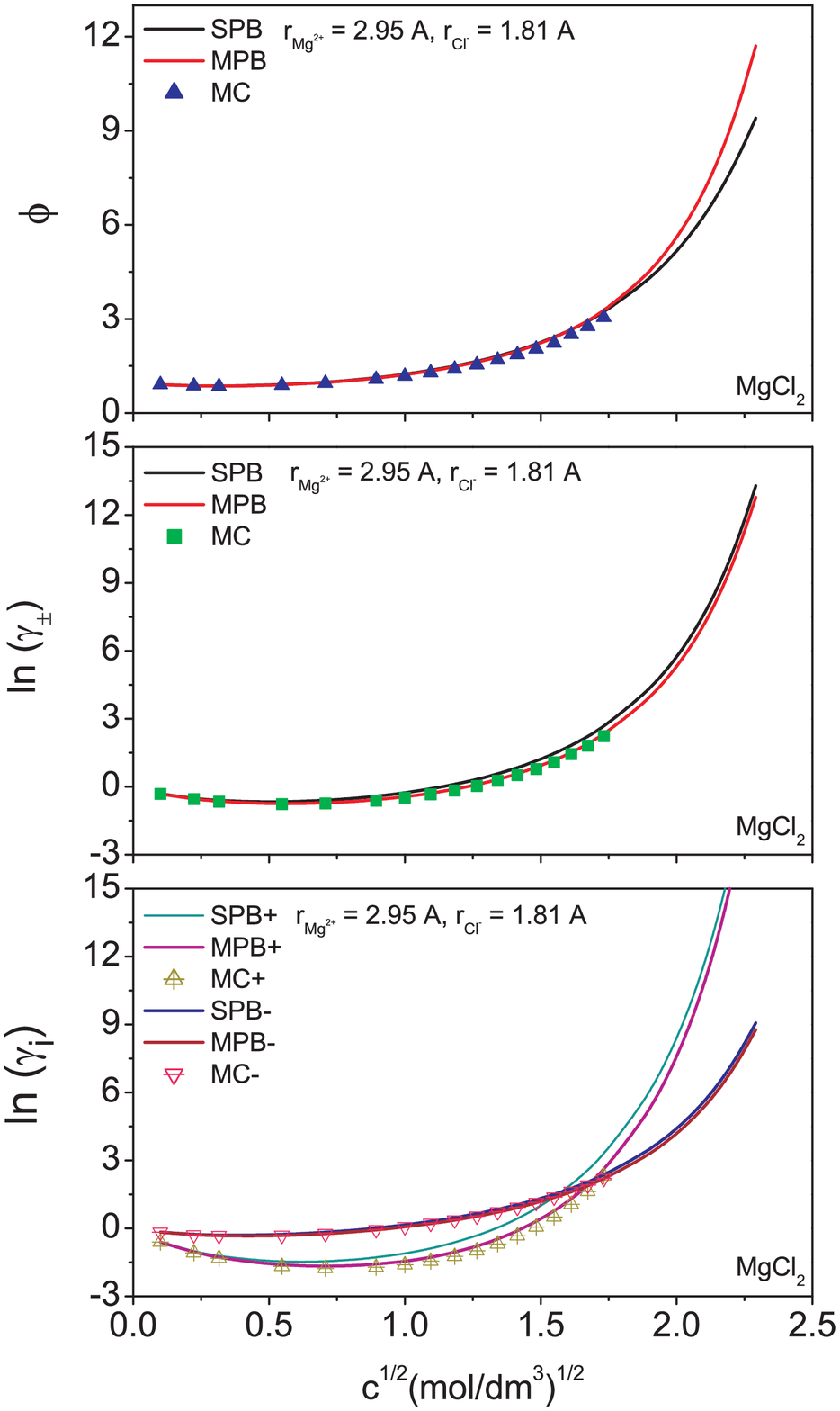}
\end{center}
\end{minipage}
\begin{minipage}{0.5\textwidth}
\begin{center}
\includegraphics[width=0.99\textwidth]{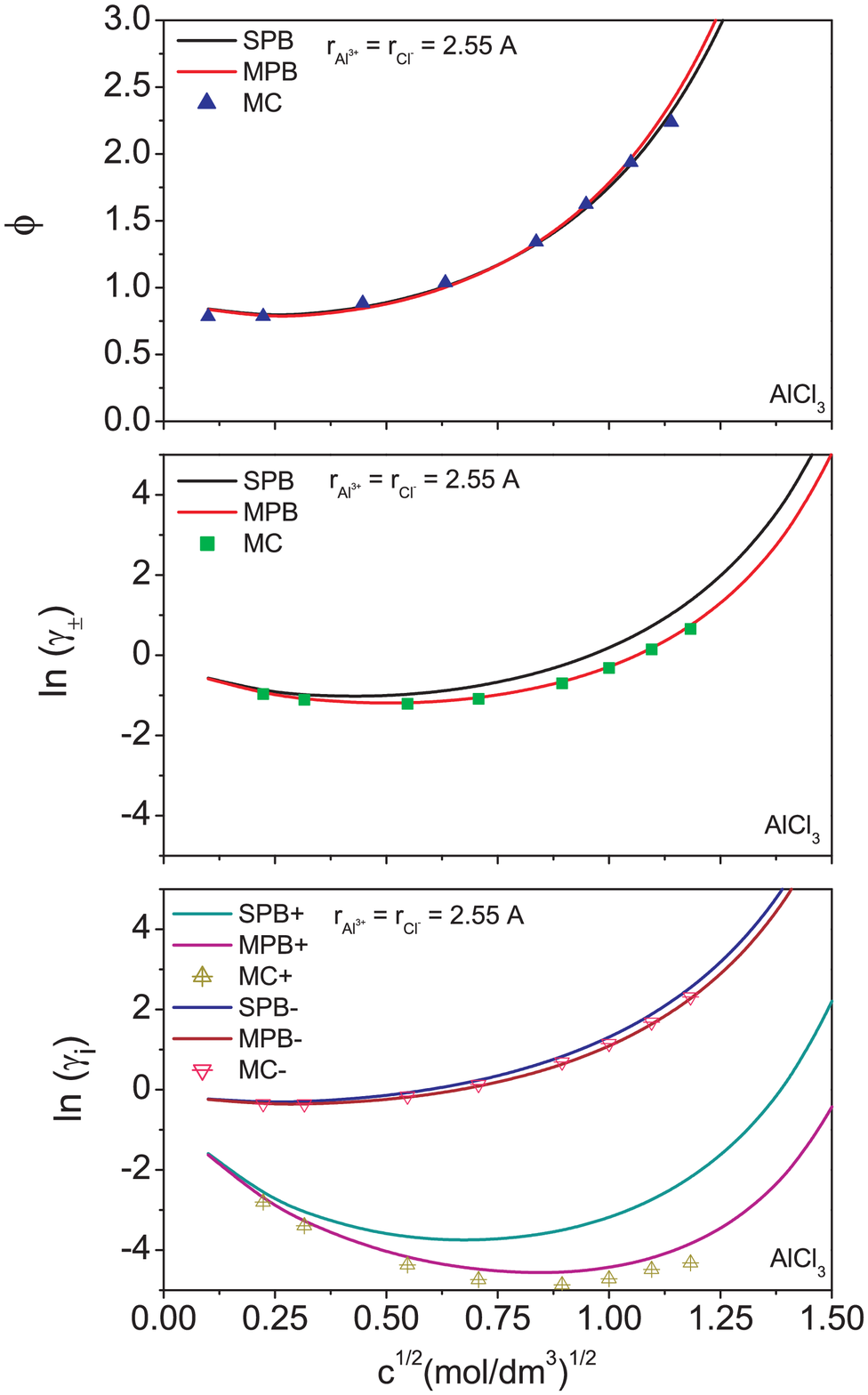}
\end{center}
\end{minipage}
\begin{minipage}{0.49\textwidth}
\begin{center}
\caption{(Colour online) MC and theoretical (SPB and MPB) osmotic coefficient (upper panel),
	natural logarithm of the mean activity coefficient (middle panel), and natural logarithm of the individual activity coefficients for MgCl$_{2}$ using PM (lower panel). The MC data are from references \cite{abbas1,abbas2}.}
\label{figure-7}
\end{center}
\end{minipage}
\begin{minipage}{0.49\textwidth}
\begin{center}
\caption{(Colour online) MC and theoretical (SPB and MPB) osmotic coefficient (upper panel),
	natural logarithm of the mean activity coefficient (middle panel), and natural logarithm of the individual activity coefficients for AlCl$_{3}$ using RPM (lower panel). The MC data are from references \cite{abbas1,abbas2}.}
\label{figure-8}
\end{center}
\end{minipage}
\vspace{-2mm}
\end{figure}


\begin{figure}[!t]
\centerline{\includegraphics[width=0.49\textwidth]{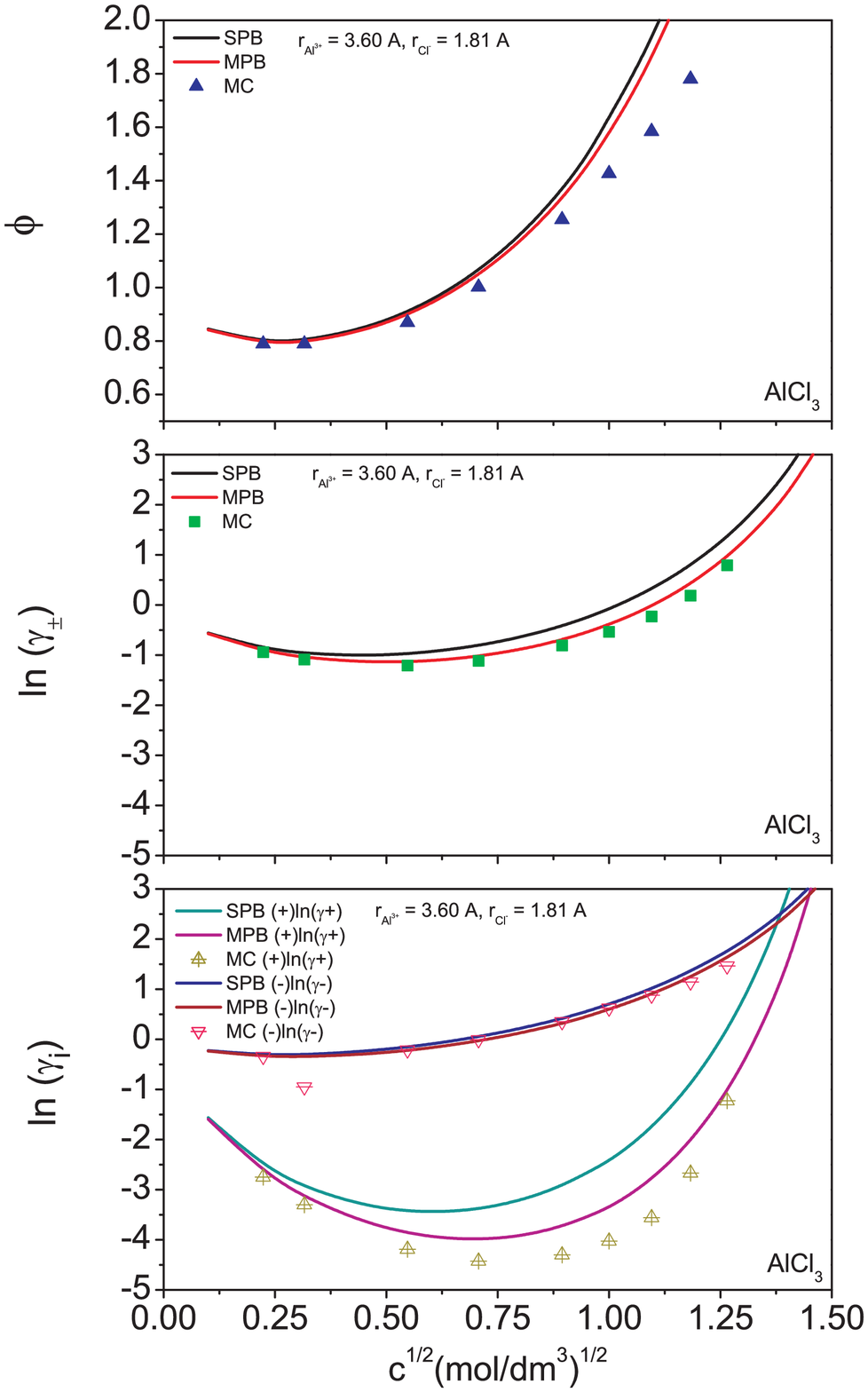}}
\caption{(Colour online) MC and theoretical (SPB and MPB) osmotic coefficient (upper panel),
	natural logarithm of the mean activity coefficient (middle panel), and natural logarithm of the individual activity coefficients for AlCl$_{3}$ using PM (lower panel). The MC data are from references \cite{abbas1,abbas2}.}
\label{figure-9}
\end{figure}

Abbas et al. \cite{abbas1,abbas2} simulated osmotic coefficients, individual and
mean activity coefficients of a total of 104 salt solutions. We have utilized their MC parameters
to obtain the SPB and MPB results for the same systems, which encompass a wide range of concentration,
ionic sizes, and valence combinations of 1:1, 2:1, and 3:1 \cite{quinones}.

We will present here, a sampling of the results for 1:1, 2:1, and 3:1 valence systems, which will
involve both the RPM and PM. For many of the electrolyte systems, Abbas et al.\cite{abbas1,abbas2}
obtained a single common ionic radius optimized through comparing the simulated and experimental
osmotic coefficient. We have chosen such systems because this affords a comparison of the RPM and PM for
the same experimental system. For the 1:1 case, the results for LiOH salt are shown for the RPM (figure~\ref{figure-4}) and PM
(figure~\ref{figure-5}). LiOH is a widely used salt in rechargeable batteries and hence
the importance of its thermodynamic properties. It is noted that for the symmetric valence RPM
situation in figure~\ref{figure-4}, we have $\ln(\gamma _{+})=\ln(\gamma _{-})=\ln(\gamma _{\pm})$. In both RPM
and PM cases, the MPB $\phi $, $\ln(\gamma _{i})$, and $\ln (\gamma _{\pm})$ are almost quantitative
with the simulations, while not unexpectedly, the SPB values are qualitative and deviating at higher
concentrations.

For the 2:1 case, we have chosen the MgCl$_{2}$ salt system, with the RPM and PM results
being displayed in figure~\ref{figure-6} and figure~\ref{figure-7}, respectively. Again, the MPB remains nearly quantitative,
and the SPB is qualitative but close to the simulations.

For a still higher asymmetric case of a 3:1 system, we have the AlCl$_{3}$ electrolyte.
The RPM results for this system are in figure~\ref{figure-8} and the PM results are in figure~\ref{figure-9}. In these figures,
the behaviour of the curves seen for the 1:1 and 2:1 cases continues, although for this high
valence asymmetry situation, even the MPB reveals some deviations from the MC.
The deviations shown by the SPB are bigger although the theory is still qualitative.

The behaviour pattern seen in these figures for the SPB and MPB curves vis-a-vis the
simulations is repeated for the remaining 95(104--9) cases not shown here, the most conspicuous
general feature being the consistent overall agreement between the theories and the MC results for
almost all the simulations, this being true for both the individual and mean activity coefficients.

\vspace{-3mm}
\section{Conclusions}

We have employed the symmetric Poisson-Boltzmann and the
modified Poisson-Boltzmann theories of statistical mechanics to characterize
the thermodynamics of electrolyte solutions. The osmotic coefficient, the
individual activity coefficients and hence the mean activity coefficients
of 104 primitive model electrolyte systems with arbitrary ionic sizes and
ionic valences were calculated and the results were compared with the corresponding
Monte Carlo simulations. In addition, the theoretical predictions were also
contrasted with the experimentally measured activity coefficients of HCl and
NaCl solutions.

We have found that overall the SPB and MPB theories reproduce the MC simulation
results to a remarkable accuracy for 1:1 electrolytes, and to a slightly lesser extent
for the asymmetric 1:2 and 1:3 valence systems. At concentrations higher than approximately
$2$ mol/dm$^3$, discrepancies between SPB and MPB tend to become relatively more prominent
even for 1:1 valence cases. This can be clearly appreciated in every graph of MC vs MPB and
SPB and is rooted in the neglect of interionic correlations in the classical mean field
approximation \cite{kirkwood}. Although at dilute solution concentrations, the influence of correlations
is minimal, they can be substantial at higher concentrations. These findings lead to the main
conclusion of this work: The agreement of the MC data to the SPB, MPB results is excellent at small
concentrations, whereas the SPB and MPB theories tend to deviate from each other at higher
concentrations. Overall, the MPB reproduces the MC results semi-quantitatively or better.

The comparison of the experimental individual activity coefficients of HCl solutions with
the corresponding simulations and the theoretical results reveals some shortcomings of the
primitive model itself. Indeed, the discrepancy between the MC and the experiments must be
due to the inadequacy of the model. However, the consistency of the mean activity coefficients
implies some cancellation of errors.

Perhaps the most serious approximation in the PM is the continuum
approximation of the solvent, this limitation having long been known \cite{robinson}.
Besides steric effects, a dissimilar polarization of the water molecules in  different
ion hydration shells would be expected to have a critical bearing on the individual
activity coefficients. The classical Born solvation model uses the continuum solvent,
so it cannot give the necessary molecular description. Born models are still useful though,
a fairly recent example being the use of simulated PM results with a Born description of the
ion-water interaction \cite{valisko}.

The 104 electrolyte solutions, for which MC data are available, cover a wide range of
concentration, ion sizes and valence combinations. The agreement of the MPB predictions, and
to a lesser extent the agreement of the SPB predictions \cite{quinones} with the relevant
simulation data, suggest the viability of these theoretical approaches in describing the
thermodynamics of charged fluids in the electrolyte solution regime within the framework
of the primitive model.

\vspace{-1mm}
\section*{Supplementary information}

The results for the SPB and MPB thermodynamics for the 104 electrolyte systems
have been archived at the Institutional Repository of the University of Puerto Rico. These can be
accessed at  \url{https://dire.upr.edu/handle/11721/1698} web site [click ``Thesis Part2 (Appendix)'' on the
left-hand side of the screen].

\vspace{-1mm}
\section*{Acknowledgements}

We would like to thank Ms. Gladys D\'{\i}az V\'{a}zquez for her help during the early
stages of this project with some of the numerical analysis of the theoretical results. We also thank
Dr. Z. Abbas of the University of Gothenburg for sending us the numerical values of their Monte Carlo
data for 104 electrolytes.

\newpage

\ukrainianpart

\title[]{Термодинаміка  примітивної моделі електролітів у симетричній і модифікованій теоріях Пуассона-Больцмана. Порівняльне дослідження із симуляціями Монте Карло}

\author{A.O. Квінонес\refaddr{label1}, Л.Б. Буян\refaddr{label1}, К.В. Оусвейт\refaddr{label2}}

\addresses{
	\addr{label1} Лабораторія теоретичної фізики, факультет фізики, Університет Пуерто Ріко,\\ Пуерто Ріко 00925-2537, США
	\addr{label2} Факультет прикладної математики, Університет Шеффілда,
	Шеффілд S3 7RH, Великобританія
}

\makeukrtitle

\begin{abstract}
	Обчислено осмотичні коефіцієнти, індивідуальні та середні коефіцієнти  активності примітивної моделі розчинів електролітів при різних молярних концентраціях з використанням симетричної і модифікованої теорій Пуассона-Больцмана. 
	Теоретичні результати порівнюються з широким спектром даних симуляцій Монте Карло, отриманих 
	Абасом та ін. [Fluid Phase Equilib., 2007, 
	\textbf{260}, 233; J. Phys. Chem. B, 2009, \textbf{113}, 5905]. Узгодження між передбаченнями модифікованої теорії Пуассона-Больцмана  та ``\emph{точними}'' результатами симуляцій є майже кількісним для моновалентних солей, будучи напів-кількісним чи   кращим для більш високих і багатовалентних солей. З іншого боку, результати симетричної  теорії Пуассона-Больцмана  є дуже добрими для моновалентних солей, але мають тенденцію відхилятися при вищих концентраціях і/або для багатовалентних систем. Також здійснено порівняння деяких нещодавніх експериментальних значень для коефіцієнтів активності розчинів  HCl (індивідуальних і середніх активностей) і розчинів  NaCl  (тільки середня активність) із симетричною та модифікованою теоріями Пуассона-Больцмана, а також з Монте Карло симуляціями.
	
	\keywords електроліти, примітивна модель, симетрична теорія Пуассона-Больцмана, модифікована теорія Пуассона-Больцмана, Монте Карло симуляції
	
\end{abstract}

\end{document}